\documentclass[aps,prl,floatfix,twocolumn,showpacs,nofitinbib,hyperref=pdftex,citeautoscript]{revtex4-1}
\usepackage{epsfig}
\usepackage{amsmath}
\usepackage{amssymb}
\usepackage{amsfonts}
\usepackage{ dsfont }
\usepackage{ comment,color }
\usepackage{lipsum}
\usepackage{hyperref}

\hypersetup{colorlinks=true,linkcolor=blue,anchorcolor=blue,citecolor=blue,filecolor=blue,urlcolor=blue,bookmarksnumbered=true,pdfview=FitB}

\newcommand{\bk}{{{\bf{k}}}}

\newcommand{\bq}{{\bf{q}}}

\newcommand{\beqa}{\begin{eqnarray}}
\newcommand{\eeqa}{\end{eqnarray}}

\newcommand{\ua}{\uparrow}
\newcommand{\da}{\downarrow}

\begin{document}

\hsize\textwidth\columnwidth\hsize\csname@twocolumnfalse\endcsname

\title{Majorana Superconducting Qubit}
\author{Constantin Schrade and Liang Fu}
\affiliation{Department of Physics, Massachusetts Institute of Technology, 77 Massachusetts Ave., Cambridge, MA 02139}

\date{\today}

\vskip1.5truecm
\begin{abstract}
We propose a platform for universal quantum computation that uses conventional $s$-wave superconducting leads to address a topological qubit stored in spatially separated Majorana bound states in a multi-terminal topological superconductor island. Both the manipulation and read-out of this ``Majorana superconducting qubit'' are realized by tunnel couplings between Majorana bound states and the superconducting leads. The ability of turning on and off tunnel couplings on-demand by local gates enables individual qubit addressability while avoiding cross-talk errors. By combining the scalability of superconducting qubit and the robustness of topological qubits, the Majorana superconducting qubit may provide a promising and realistic route towards quantum computation.
\end{abstract}

% 03.67.Lx: Quantum computation architectures and implementations
% 74.50.+r: Tunneling phenomena; Josephson effects
% 85.25.Cp Josephson devices
% 71.10.Pm: Fermions in reduced dimensions

\maketitle
Superconducting circuits are among the leading platforms for quantum computing.
Their main building block is the superconducting qubit which is based on the Josephson tunnel junction, a non-dissipative and non-linear electrical element that enables long-coherence times \cite{bib:Paik2011,bib:Rigetti2012,bib:Chang2013} and high-fidelity gate operations \cite{bib:Barends2014,bib:Sheldon2016}. With recent advances in scaling to qubit arrays and surface code architectures
\cite{bib:Helmer2009,bib:DiVincenzo2009,bib:Fowler2012,bib:Corcoles2015,bib:Kelly2015,bib:Takita2016}, significant efforts are being made to minimize errors due to unintentional cross-talk between qubits \cite{bib:Gambetta2012,bib:Saira2014,bib:Kelly2015,bib:Blumoff2016} and to avoid leakage into non-computational states \cite{bib:Kringhoj2017, bib:Hutchings2017}.

\begin{figure}[!t] \centering
\includegraphics[width=1\linewidth] {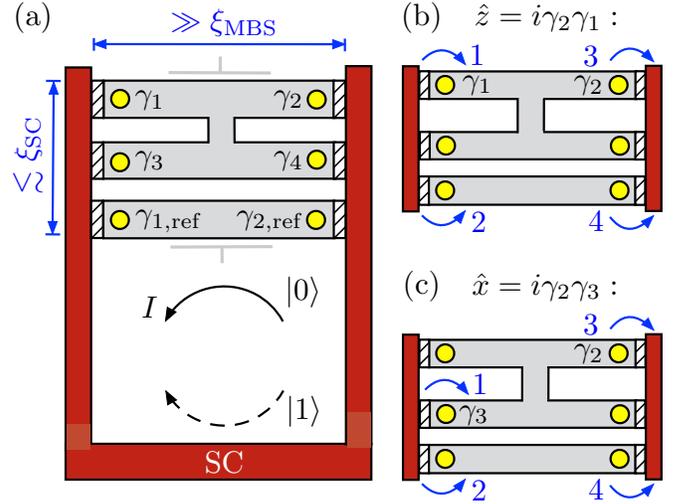}
\caption{(Color online)
(a) Minimal setup for a MSQ experiment. A four-terminal TSC island realizing a single MSQ and a two-terminal reference island (both gray) are placed in an $s$-wave SC Josephson junction (red). The horizontal extent of the islands are assumed to be larger than the localization length $\xi_{\text{MBS}}$ of the MBSs $\gamma_{\ell}$ and $\gamma_{\ell,\text{ref}}$ (yellow) which emerge at the terminal points of the islands. The vertical extent of the unit cell is assumed to be at most of the order of the superconducting coherence length $\xi_{\text{SC}}$ thereby enabling Cooper pair splitting between the superconducting leads mediated by the MBSs. With the suitable choice of tunnel couplings discussed in the text, the states of the MSQ, $|0\rangle$ and $|1\rangle$, can be read-out be measuring the direction of the persistent Josephson current in a loop.
(b) Typical Cooper pair splitting process between one of the four-terminal islands and the two-terminal reference islands utilized for implementing rotations around the $z$-axis of the MSQ Bloch sphere.
(c) Same as (a) but for rotations around the $x$-axis of the MSQ Bloch sphere
}\label{fig:1}
\end{figure}

In this work, we introduce a new platform for universal quantum computing that combines the scalability of the superconducting qubit and the robustness of Majorana qubit. The key element in our proposal is a multi-terminal topological superconductor (TSC) island with spatially separated Majorana bound states (MBSs), used as a weak link between superconducting electrodes. The minimal setup is a Josephson junction that consists of two TSC weak links in parallel within a superconducting loop, as shown in Fig.~\ref{fig:1}(a). Both TSC islands operate in the Coulomb blockade regime and mediate the Josepson coupling via virtual charge fluctuations.
The first island hosts four MBSs ($\gamma_1, \gamma_2, \gamma_3, \gamma_4$) at the four terminals, which stores a single topological qubit.  The second is a two-terminal island with two MBSs ($\gamma_{1,\text{ref}}, \gamma_{2,\text{ref}}$) used for qubit manipulation and readout only.
The full set of single-qubit rotations is achieved by selectively turning on and off the tunnel couplings between individual MBSs and the SC electrodes that enable different Cooper pair splitting processes, see Fig.~\ref{fig:1}(b) and (c). The qubit read-out is achieved by measuring the persistent supercurrent in the loop, see Fig.~\ref{fig:1}(a). We term this basic building block---Majorana-based qubit in an all-superconducting circuit---``Majorana superconducting qubit'' (MSQ).

Compared to the conventional superconducting qubit, the MSQ is expected to have several advantages. First, the nonlocal storage of quantum information in well-separated MBSs makes the MSQ protected from decoherence under local perturbations at a physical level \cite{bib:Kitaev}. The MSQ is also insensitive to global electrostatic fluctuations that couple to the total charge on the TSC island \cite{bib:Vijay2016_2, bib:Karzig2016, bib:Plugge2017}.
Second, since a MSQ is formed by two {\it topologically degenerate} states that are separated from the excited states by the TSC gap, leakage to non-computational states, which is a common problem encountered in gate operations on weakly-anharmonic transmon qubits, is strongly suppressed.
Third, both gate operations and qubit read-out are realized solely by tuning tunnel couplings between the TSC island and the superconducting leads, which can be turned on and off on-demand through local gates as recently demonstrated in semiconductor based superconducting qubits \cite{bib:Lange2015,bib:Larsen2015,bib:Casparis2017}.
Importantly, a specific set of tunnel couplings are to be turned on only during the gate operation and measurement. The ability of pinching off unwanted tunnel couplings allows us to address MSQ individually without cross-talk errors. This provides an advantage over flux-controlled tuning of Josephson energy in transmon and hybrid transmon-Majorana qubits \cite{bib:Hyart2013}.

The use of superconducting interference effect for qubit manipulation and read-out in our proposal constitutes a key advantage over recently proposed Majorana platforms for quantum computation \cite{bib:Vijay2016_2, bib:Karzig2016, bib:Plugge2017}, where MBSs are addressed by Aharonov-Bohm interference of single electrons \cite{bib:Fu2010}. The latter requires electron phase coherence in a non-superconducting lead. The limited phase coherence length in InAs nanowires \cite{bib:Liang2009,bib:Doh2009} places an important constraint on device geometries. In contrast, in our setup, there is no upper bound on the size of the superconducting loop, as the persistent supercurrent is dissipationless. Importantly, the separation between the two parallel TSC islands is required to be shorter than the superconducting coherence length, in order to enable Cooper pair splitting processes. For conventional superconductors such as aluminium, the coherence length can be several hundreds of nanometers \cite{bib:Cyrot1992}.

{\it Setup.}
The setup for a minimal MSQ experiment enabling both single-qubit rotations and read-out is depicted in Fig.~\ref{fig:1}(a).
It comprises a single four-terminal islands as well as a two-terminal reference island. The MBSs which form at the terminal points $\ell$ are denoted by $\gamma_{\ell}$ for the four-terminal island and by $\gamma_{\ell,\text{ref}}$ for the two-terminal reference island. We assume that the horizontal extent of the islands is much larger than the MBS localization length $\xi_{\text{MBS}}$ such that the wavefunction hybridization of MBSs localized at opposite terminals is negligible and, therefore, all MBSs reside at zero energy. Since the TSC islands are of mesoscopic size, each island acquires a finite charging energy
\begin{subequations}
\begin{align}
U &= (ne-Q)^{2}/2C,
\\
U_{\text{ref}} &= (n_{\text{ref}}e-Q_{\text{ref}})^{2}/2C_{\text{ref}}.
\end{align}
\end{subequations}
Here, $n$ and $n_{\text{ref}}$ denote the number of unit charges on the islands. Furthermore, $Q$ and $Q_{\text{ref}}$ are gate charges which are continuously tunable via gate voltages across capacitors with capacitances $C$ and $C_{\text{ref}}$, respectively. For simplicity, we will focus on the case of equal capacitances, $C=C_{\text{ref}}$. Assuming the strong Coulomb blockade regime and a tuning of the gate charges $Q$, $Q_{\text{ref}}$ close to integer values, the total fermion parities of the islands obey the constraints \cite{bib:Fu2010,bib:Xu2010},
\begin{subequations}
\begin{align}
\gamma_{1}\gamma_{2}\gamma_{3}\gamma_{4} &= (-1)^{n_{0}+1}, \label{Constraint1}
\\
i\gamma_{1,\text{ref}}\gamma_{2,\text{ref}}&= (-1)^{n_{0,\text{ref}}}. \label{Constraint2}
\end{align}
\end{subequations}
In writing down these expressions, we have omitted finite-energy quasiparticle contributions, which is a justified provided that the island energy gaps define the largest energy scale of the setup. A consequence of the constraints given in both
Eq.~\eqref{Constraint1} and Eq.~\eqref{Constraint2} is that the dimensionality of the ground state subspace at zero charging energy decreases by a factor of two for all islands. In particular, for the four-terminal island, the four-fold degenerate ground state subspace at zero charging energy reduces to a two-fold degenerate ground state subspace which makes up the MSQ. The Pauli matrices acting on the each of the two MSQs are given by
\begin{equation}
\begin{split}
\hat{x}= i\gamma_{2}\gamma_{3}, \quad \hat{y}= i\gamma_{1}\gamma_{3}, \quad
\hat{z}= i\gamma_{2}\gamma_{1}.
\end{split}
\end{equation}

As depicted in Fig.~\ref{fig:1}(a), the TSC islands are placed in a Josephson junction of two bulk, $s$-wave superconducting leads
that are labelled by $m=\text{L,R}$ and are used to address the MSQs through tunable tunnel couplings. The BCS (Bardeen-Cooper-Schrieffer) Hamiltonian of the superconducting leads is given by,
\begin{equation}
\label{BCS}
H_{0}=\sum_{m=\text{L,R}}\sum_{\bk} \Psi_{m,\bk}^\dagger \left(
\xi_{\bk}\eta_{z}+\Delta_{m}\eta_{x}e^{i\varphi_{m}\eta_{z}}
\right)\Psi_{m,\bk},
\end{equation}
where $\Psi_{m,\bk}=(c_{m,\bk\ua},c^{\dag}_{m,-\bk\da})^{T}$ denotes a Nambu spinor with
$c_{m,\bk s}$ being the annihilation operator of an electron with momentum $\bk$ and Kramers index $s=\ua,\da$.
The magnitude and phase of the superconducting ordering parameter are given by $\Delta_{m}$ and $\varphi_{m}$, respectively. The Pauli matrices $\eta_{x,y,z}$ are acting in Nambu space. For simplicity, we will assume that the magnitudes of the SC order parameters are identical for both leads, $\Delta\equiv\Delta_{m}$.

The tunneling Hamiltonians which couple the SC leads to the MBSs at the terminal points are given by
\begin{subequations}
\begin{align}
H_{T}
&=
\sum_{m,\ell}
\sum_{\bk,s}
\lambda^{s}_{m\ell} \
c^{\dag}_{m,\bk s}
\gamma_{\ell}
e^{-i\phi/2}
+
\text{H.c.},
\\
H_{T,\text{ref}}
&=
\sum_{m,\ell}
\sum_{\bk,s}
\lambda^{s}_{m\ell,\text{ref}} \
c^{\dag}_{m,\bk s}
\gamma_{\ell,\text{ref}}
e^{-i\phi_{\text{ref}}/2}
+
\text{H.c.},
\end{align}
\end{subequations}
for the four-terminal and the two-terminal reference islands, respectively. For simplicity, the tunnel couplings are taken to be  point-like. This is justified provided that the separation between individual tunneling contacts is much smaller than the superconducting coherence length $\xi_{\text{SC}}$.
In the subsequent discussions, we will assume that the lead electrons will only couple to nearby MBSs, \textit{i.e.}, $\lambda^{s}_{\text{L}2}=\lambda^{s}_{\text{L}4}=\lambda^{s}_{\text{R}1}=\lambda^{s}_{\text{R}3}=0$ and $\lambda^{s}_{\text{L}2,\text{ref}}=\lambda^{s}_{\text{R}1,\text{ref}}=0$. This is justified if the MBS localization length $\xi_{\text{MBS}}$ is much larger than horizontal segments of the islands. The remaining non-zero tunnel couplings are assumed to take on the most general complex and spin-dependent form. Moreover, the operator $e^{\pm i\phi/2}$ and $e^{\pm i\phi_{\text{ref}}/2}$  increase/decrease the total charge of the four-terminal island or the two-terminal reference island
by one charge unit, $[n,e^{\pm i\phi/2}]=\pm e^{\pm i\phi/2}$ and $[n_{\text{ref}},e^{\pm i\phi_{\text{ref}}/2}]=\pm e^{\pm i\phi_{\text{ref}}/2}$, while the MBSs operators $\gamma_{\ell}$ and $\gamma_{\ell,\text{ref}}$ change the electron number parity of respective islands \cite{bib:Fu2010}. In summary, the Hamiltonian for a minimal MSQ experiment is given by
$
H=H_{0}+U+U_{\text{ref}}+H_{T}+H_{T,\text{ref}}.
$

{\it Single-qubit control.}
In this section, we describe the simplest MSQ experiments which allows for both read-out and manipulation of a single MSQ. In combination with the two-qubit entangling operation introduced in the next section, this will enable universal quantum computation \cite{bib:Brylinski2001}.

As a starting point, we discuss rotations around the $z$-axis of the MSQ Bloch sphere as well as the read-out of the $\hat{z}$-eigenvalue. We, therefore, consider the case when only the couplings to the two-terminal reference island and the two couplings $\lambda^{s}_{\text{L}1}$ and $\lambda^{s}_{\text{R}2}$ at opposite boundaries of the four-terminal island are non-vanishing, see Fig.~\ref{fig:1}(b).

In this case, the Josephson coupling between the SC leads is mediated exclusively by fourth-order co-tunnelling processes via both the two-terminal and the four-terminal island. An example of such a fourth-order process involves extracting two electrons which form a Cooper pair from one of the SC leads and placing them onto the two spatially separated islands in the first two intermediate steps. Such a coherent splitting of Cooper pairs requires the vertical distance of the islands to be smaller than the superconducting coherence length $\xi_{\text{SC}}$ and leads to virtually excited states of order $U\equiv e^{2}/2C$ on both islands. In the final two intermediate steps, the Cooper pair is recombined on the other lead, and the system thereby returns to its ground state. It is crucial to highlight that no direct Cooper pair tunnelling via a single island can happen due to the conflicting pairing symmetries of islands and leads \cite{bib:Zazunov2012,bib:Zazunov2017,bib:Zazunov2018}. Additionally, second-order co-tunnelling processes occurring separately between each SC lead and the islands are suppressed by a large charging energy $U$ \cite{bib:Schrade2018}.

The amplitudes of all Cooper pair splitting processes can be computed perturbatively in the weak-tunnelling limit, $\pi\nu_{m}|\lambda^{s}_{m\ell,\text{ref}}\lambda^{s'}_{m\ell}|\ll\Delta,U$ with $\nu_{m}$ the normal-state density of states per spin of the lead $m$ at the Fermi energy. The results are summarized by an effective Hamiltonian acting on the BCS ground states of the leads and the charge ground states of the islands \cite{bib:supplemental},
\begin{equation}
H_{z,\text{eff}} = (-1)^{n_{0,\text{ref}}+1} (J_{12}+\tilde{J}_{12})  \cos(\varphi+\varphi_{12}) \hat{z}.
\label{Heffz}
\end{equation}
Here, we have introduced the couplings constants and the anomalous phase shift,
\begin{align}
J_{\ell\ell'} &=
\frac{32|\Gamma_{\text{L}\ell}\Gamma_{\text{R}\ell'}|
}{\pi^{2}\Delta}
\int^{\infty}_{1}
\frac
{
\mathrm{d}x \text{ }�� \mathrm{d}y
}{
f(x)f(y)
\left[f(x)+f(y)\right]g(x)g(y)},\nonumber
\\
\tilde{J}_{\ell\ell'} &=
\frac{64|\Gamma_{\text{L}\ell}\Gamma_{\text{R}\ell'}|
}{\pi^{2}\Delta}
\int^{\infty}_{1}
\frac
{
\mathrm{d}x \text{ }�� \mathrm{d}y
}{
f(x)f(y)
\left[g(x)+g(y)\right]g(x)g(y)},\nonumber
\\
\varphi_{\ell\ell'}&=
\arg[
\Gamma^{*}_{\text{L}\ell}\Gamma_{\text{R}\ell'}
].
\label{Couplings}
\end{align}
Moreover, we have defined the functions $f(x)\equiv\sqrt{1+x^{2}}$, $g(x)\equiv\sqrt{1+x^{2}}+U/\Delta$ as well as the hybridization
\begin{equation}
\Gamma_{m\ell}
\equiv
\pi\nu_{m}
(
\lambda^{\da}_{m\ell,\text{ref}}\lambda^{\ua}_{m\ell}
-
\lambda^{\ua}_{m\ell,\text{ref}}\lambda^{\da}_{m\ell}
).
\label{Condition}
\end{equation}
The effective Hamiltonian given in Eq.~\eqref{Heffz} is the first main finding of our work. Three aspects are noteworthy:

(1) The unitary time-evolution operator of the effective Hamiltonian implements rotations around the $z$-axis of the MSQ Bloch sphere. More explicitly, by pulsing the couplings and phases of the effective Hamiltonian for a time $t_{z}$ such that $(-1)^{n_{0,\text{ref}}+1} \int^{t_{z}}[J_{12}(t)+\tilde{J}_{12}(t)]  \cos[\varphi(t)+\varphi_{12}(t)]=\hbar\theta_{z}/2$ a rotation by an arbitrary angle $\theta_{z}$ around the $z$-axis of the MSQ Bloch sphere is achieved.

(2) A choice of basis for the MSQ is given by the eigenstates of the $\hat{z}$-Pauli operator. Thus, a read-out of the MSQ in this basis amounts to measuring the eigenvalues $z=\pm1$ of the $\hat{z}$-Pauli operator. This can be accomplished by measuring the sign of the resulting zero-temperature Josephson current,
\begin{equation}
I=\frac{2e}{\hbar}(-1)^{n_{0,\text{ref}}} (J_{12}+\tilde{J}_{12})  \sin(\varphi+\varphi_{12}) z.
\end{equation}
For $n_{0,\text{ref}}$ being odd (even), a negative (positive) critical current implies that $z=+1$ while a positive (negative)
critical currents implies that $z=-1$, see Fig.~\ref{fig:1}(a).

(3)  A necessary requirement for a non-zero effective Hamiltonian is that $\Gamma_{1\text{L}}\neq0$ and $\Gamma_{2\text{R}}\neq0$. These conditions are in general fulfilled granted that the MBSs in the two islands couple asymmetrically to the two spin-species of the SC leads, see Eq.~\eqref{Condition}. In fact, the strength of the Josephson coupling is maximized if the MBSs in different islands couple to opposite spin species in the SC leads. When the horizontal island segments are realized by parallel semiconductor nanowires \cite{bib:Mourik2012,bib:Das2012,bib: Churchill2013,bib:Gazibegovic2017,bib:Vaitiekenas2018,bib:Krizek2018} coupled by a SC bridge \cite{bib:Plugge2017,bib:Karzig2016}, there are multiple ways on how the desired asymmetry can be accomplished: One option is to have a common spin polarization in the two nanowires and a finite spin-orbit coupling in the tunnelling barriers. As a result of the finite spin-orbit coupling the spin of the lead electrons will perform rotations with a period of $\pi\hbar^{2}/\alpha m^{*}$ inside the tunnelling barrier \cite{bib:Bercioux2015}. Here, $\alpha$ denotes the strength of the Rashba spin-orbit coupling in the tunnelling barrier and $m^{*}$ is the effective electron mass. Thus by appropriately adjusting the length of the tunnelling barriers, we can transport a Cooper pair across the junction by pure spin-flip tunnelling in the barriers to the reference island and pure normal tunnelling in the barriers to the four-terminal island. An alternative option is to generate different (ideally opposite) spin polarization in the two nanowires by using local magnetic fields. Such fields could be obtained by coating the wires with ferromagnets that produce different exchange fields.

So far, we have focused on rotations around the $z$-axis of the MSQ Bloch sphere. We will now show that rotations around the $x$-axis can be realized similarly. To this end, we choose  $\lambda^{s}_{3\text{L}}$, $\lambda^{s}_{2\text{R}}$ , $\lambda^{s}_{1,\text{ref}}$ and $\lambda^{s}_{2,\text{ref}}$ as the only non-zero tunnel couplings, see Fig.~\ref{fig:1}(c). The Josephson coupling between the superconducting leads is again facilitated solely by Cooper pair splitting processes via the TSC islands. In the weak tunnelling limit, the amplitudes of these processes are summarized by an effective Hamiltonian acting on the BCS ground states of the leads and the charge ground states of the islands \cite{bib:supplemental},
 \begin{equation}
H_{x,\text{eff}} = (-1)^{n_{0,\text{ref}}+1} (J_{32}+\tilde{J}_{32})  \cos(\varphi+\varphi_{32}) \hat{x},
\label{Heffx}
\end{equation}
It is not hard to see that pulsing the couplings and phases of this effective Hamiltonian for a time  $t_{x}$ such that $(-1)^{n_{0,\text{ref}}+1} \int^{t_{z}}[J_{32}(t)+\tilde{J}_{32}(t)]  \cos[\varphi(t)+\varphi_{32}(t)]=\hbar\theta_{x}/2$
enables rotations by an angle $\theta_{x}$ around the $x$-axis of the MSQ Bloch sphere. Combining this observation with the results of Eq.~\eqref{Heffz} allows us to perform rotations around two independent axes on the Bloch sphere and, therefore, enables the implementation of arbitrary single-qubit gates acting on the MSQ.

{\it Two-qubit gates.}
\begin{figure}[!t] \centering
\includegraphics[width=1\linewidth] {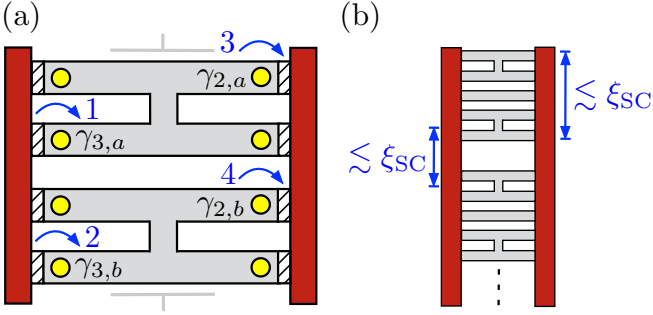}
\caption{(Color online)
(a) Typical Cooper pair splitting process between two four-terminal islands implementing a two-qubit entangling $XX_{\chi}\equiv\exp(-i\chi \hat{x}_{a}\hat{x}_{b})$ gate for some parameter $\chi$.
(b) Linear array of unit cells. The separation between unit cells is assumed to be smaller than the superconducting coherence length $\xi_{\text{SC}}$ which allows for coherent exchange of quantum information by SWAP gates between neighbouring cells.
}\label{fig:2}
\end{figure}
What remains to be shown to achieve universality in our setup is the implementation of a two-qubit entangling gate.
This will be the topic of the present section. As a starting point, we consider two four-terminal islands labelled by $j=a,b$ and choose $\lambda^{s}_{3\text{L},a}$, $\lambda^{s}_{2\text{R},a}$, $\lambda^{s}_{3\text{L},b}$, $\lambda^{s}_{2\text{R},b}$ as the only non-zero tunnel couplings, see Fig.~\ref{fig:2}(a). The Cooper pair splitting processes which lead to a Josephson coupling between the superconducting leads are now entirely facilitated by the two four-terminal TSC islands. Their amplitudes can be computed in the weak-tunnelling limit, $\pi\nu_{m}|\lambda^{s}_{m\ell,j}\lambda^{s'}_{m\ell',j'}|\ll\Delta,U$, and are summarized by an effective Hamiltonian which acts on the BCS ground states and the charge ground states of the TSC islands \cite{bib:supplemental},
\begin{equation}
H_{\text{eff}} =  (J+\tilde{J})  \cos(\varphi+\varphi_{0}) \hat{x}_{a}\hat{x}_{b}.
\label{Heffxx}
\end{equation}
Here, we have introduced the couplings constants and the anomalous phase shift,
\begin{align}
J &=
\frac{32|\Gamma'_{\text{L}3}\Gamma'_{\text{R}2}|
}{\pi^{2}\Delta}
\int^{\infty}_{1}
\frac
{
\mathrm{d}x \text{ }�� \mathrm{d}y
}{
f(x)f(y)
\left[f(x)+f(y)\right]g(x)g(y)},\nonumber
\\
\tilde{J} &=
\frac{64|\Gamma'_{\text{L}3}\Gamma'_{\text{R}2}|
}{\pi^{2}\Delta}
\int^{\infty}_{1}
\frac
{
\mathrm{d}x \text{ }�� \mathrm{d}y
}{
f(x)f(y)
\left[g(x)+g(y)\right]g(x)g(y)},\nonumber
\\
\varphi_{0}&=
\arg[
(\Gamma'_{\text{L}3})^{*}\Gamma'_{\text{R}2}
].
\label{Couplings2}
\end{align}
Moreover, we have defined the hybridization
\begin{equation}
\Gamma'_{m\ell}
\equiv
\pi\nu_{m}
(
\lambda^{\da}_{m\ell,b}\lambda^{\ua}_{m\ell,a}
-
\lambda^{\ua}_{m\ell,b}\lambda^{\da}_{m\ell,a}
).
\label{Condition2}
\end{equation}
The effective Hamiltonian in Eq.~\eqref{Heffxx} is the second main result of our work. By pulsing the couplings and phases for a time $\tau$ such that $\int^{\tau}[J(t)+\tilde{J}(t)]  \cos[\varphi(t)+\varphi_{0}(t)]=\hbar\chi$, the unitary time-evolution operator of the effective Hamiltonian implements an $XX_{\chi}\equiv\exp(-i\chi \hat{x}_{a}\hat{x}_{b})$ gate for some parameter $\chi$. It is well-known in the literature that the $XX_{\pi/4}$-gate together single-qubit operations implements a CNOT gate \cite{bib:Maslov2017},
\begin{equation}
\text{CNOT} =
X_{-\frac{\pi}{2},b}
\cdot
Y_{-\frac{\pi}{2},a}
\cdot
X_{-\frac{\pi}{2},a}
\cdot
XX_{\frac{\pi}{4}}
\cdot
Y_{\frac{\pi}{2},a}
,
\label{CNOT}
\end{equation}
where we have introduced the single-qubit gates $X_{\theta,j}\equiv\exp(-i\theta \hat{x}_{j}/2)$ and $Y_{\theta,j}\equiv\exp(-i\theta \hat{y}_{j}/2)$ with some parameter $\theta$. We note that the CNOT-gate defined in Eq.~\eqref{CNOT} uses the MSQ $a$ as control and the MSQ $b$ as target. A CNOT$'$-gate in which the roles of control and target-qubit are reversed can readily be obtained by applying single-qubit Hadamard gates, CNOT$'=H_{a}\cdot H_{b}\cdot \text{CNOT}\cdot H_{a}\cdot H_{b}$ with $H_{\ell}=(\hat{x}_{\ell}+\hat{z}_{\ell})/\sqrt{2}$.
In conclusion, the combination of the single-qubit gates introduced in the previous section together with the two-qubit CNOT gate is sufficient for universal quantum computation with MSQs.

To assemble a scalable MSQ computer, we consider unit cells comprised of two four-terminal islands and a single reference island. This enables the implementation of a universal gate set comprised of arbitrary single-qubits gates and a two-qubit entangling gate within each unit cell. Importantly, such a unit cell can readily be scaled to a linear array of multiple unit cells as depicted in Fig.~\ref{fig:2}(b). The distance between the individual unit cells in such an array is taken to be at most of the order of the superconducting coherence length $\xi_{\text{SC}}$. The coherent exchange of quantum information between different unit cells is facilitated by SWAP gates acting on MSQs of neighbouring unit cells \cite{bib:supplemental}.

Before closing, we envision two candidate platforms for a material realization of MSQs. The first platform are parallel topologically SC nanowires \cite{bib:Mourik2012,bib:Das2012,bib: Churchill2013,bib:Gazibegovic2017,bib:Vaitiekenas2018,bib:Krizek2018} coupled via a trivially SC bridge \cite{bib:Plugge2017,bib:Karzig2016} that are placed in a Josephson junction of conventional $s$-wave SCs. Here, it is worth mentioning that Cooper pair splitting between parallel semiconducting nanowires coupled to a common superconducting electrode -- the key ingredient of our proposal -- has been observed in recent experiments \cite{bib:Baba2018}. The second platform which we envision for a MSQ realization are TSC islands defined in a heterostructure of a two-dimensional electron gas and a SC by means of top-down lithography and gating \cite{bib:Suominen2017}. A key advantage of these devices is that they may enable rapid scaling from a single MSQ to the muti-MSQ architectures of Fig.~\ref{fig:2}(b).

{\it Conclusions.}
We have put forward a platform for universal quantum computation realized by conventional $s$-wave superconducting leads addressing MSQs formed by the charge ground states of four-terminal TSC islands. We have demonstrated how single-qubit operations, qubit read-out, as well as two-qubit entangling gates are accomplished by Cooper pair spitting processes between the superconducting leads mediated by the MBSs on the islands. Since the tunnel couplings can be turned on and off on-demand, unintentional cross-talk between different MSQs is avoided. This important feature, together with the robustness of the MBSs under local environmental perturbations, may provide an alternative pathway to superconducting quantum computation.

{\it Acknowledgments.}
We would like to thank Morten Kjaergaard for helpful discussions. C.S. was supported by the Swiss SNF under Project 174980. L.F. and C.S. were supported by DOE Office of Basic Energy Sciences, Division of Materials Sciences and Engineering under Award $\text{DE-SC0010526}$.

\begin{widetext}

\newpage

\onecolumngrid

\bigskip

\begin{center}
\large{\bf Supplemental Material to `Majorana Superconducting Qubit' \\}
\end{center}
\begin{center}
Constantin Schrade and Liang Fu
\\
{\it Department of Physics, Massachusetts Institute of Technology, 77 Massachusetts Ave., Cambridge, MA 02139}
\end{center}

In the Supplemental Material, we provide more details on the derivation of the effective Hamiltonians used for the implementation of the single-qubit and the two-qubit gates.

\section{Effective Hamiltonian for the single-qubit gates}
In this first section of the Supplemental Material, we present the derivation of the effective Hamiltonians used for accomplishing single-qubit operations on the MSQ. We first consider rotations around the $z$-axis of the MSQ Bloch sphere. We, therefore, choose $\lambda^{s}_{\text{L}1}$, $\lambda^{s}_{\text{R}2}$, $\lambda^{s}_{\text{L}1,\text{ref}}$ and $\lambda^{s}_{\text{R}2,\text{ref}}$ as the only non-zero couplings connecting the superconducting leads to the TSC islands. The tunnelling Hamiltonians which were given in Eq.~(5a) and Eq.~(5b) of the main text then take on the simplified form
\begin{subequations}
\begin{align}
H_{T}
&\rightarrow
\sum_{\bk}
\left(
\lambda^{\ua}_{\text{L} 1} \
c^{\dag}_{\text{L},\bk \ua}
+
\lambda^{\da}_{\text{L} 1 } \
c^{\dag}_{\text{L},\bk \da}
\right)
\gamma_{1}
e^{-i\phi/2}
+
\left(
\lambda^{\ua}_{\text{R}2} \
c^{\dag}_{\text{R},\bk \ua}
+
\lambda^{\da}_{\text{R} 2 } \
c^{\dag}_{\text{R},\bk \da}
\right)
\gamma_{2}
e^{-i\phi/2}
+\text{H.c.},\label{TunnelingH1}
\\
H_{T,\text{ref}}
&=
\sum_{\bk}
\left(
\lambda^{\ua}_{\text{L}1,\text{ref}} \
c^{\dag}_{\text{L},\bk \ua}
+
\lambda^{\da}_{\text{L}1,\text{ref}} \
c^{\dag}_{\text{L},\bk \da}
\right)
\gamma_{1,\text{ref}}
e^{-i\phi_{\text{ref}}/2}
+
\left(
\lambda^{\ua}_{\text{R}2,\text{ref}} \
c^{\dag}_{\text{R},\bk \ua}
+
\lambda^{\da}_{\text{R}2,} \
c^{\dag}_{\text{R},\bk \da}
\right)
\gamma_{2,\text{ref}}
e^{-i\phi_{\text{ref}}/2}
+
\text{H.c.}\label{TunnelingH2}
\end{align}
\end{subequations}
As a next step, we perform a unitary rotation in the SC leads such that the MBSs in the four-terminal TSC island couple only to the spin-up lead electrons,
\begin{align}
\label{rotation}
\begin{pmatrix}
d_{\text{L},\bk \uparrow} \\
d_{\text{L},-\bk \downarrow}
\end{pmatrix}
&\equiv
\frac{\pi\nu_{\text{L}}}{
N_{\text{L}1}
}
\begin{pmatrix}
(\lambda^{\ua}_{\text{L}1})^{*} & (\lambda^{\da}_{\text{L}1})^{*} \\
-\lambda^{\da}_{\text{L}1} & \lambda^{\ua}_{\text{L}1}
\end{pmatrix}
\cdot
\begin{pmatrix}
c_{\text{L},\bk \ua} \\
c_{\text{L},-\bk \da}
\end{pmatrix}, \quad
\begin{pmatrix}
d_{\text{R},\bk \uparrow} \\
d_{\text{R},-\bk \downarrow}
\end{pmatrix}
\equiv
\frac{\pi\nu_{\text{R}}}{
N_{\text{R}2}
}
\begin{pmatrix}
(\lambda^{\ua}_{\text{R}2})^{*} & (\lambda^{\da}_{\text{R}2})^{*} \\
-\lambda^{\da}_{\text{R}2} & \lambda^{\ua}_{\text{R}2}
\end{pmatrix}
\cdot
\begin{pmatrix}
c_{\text{L},\bk \ua} \\
c_{\text{L},-\bk \da}
\end{pmatrix}.
\end{align}
Here, we have defined the normalization $N_{m\ell}\equiv\pi\nu_{m}\sqrt{
|\lambda^{\ua}_{m\ell} |^{2}+|\lambda^{\da}_{m\ell}|}$. Additionally, we remark that in the newly introduced fermionic operators $d_{m,\bk s}$ the label $s=\ua,\da$ refers to a Kramers index defined by $\mathcal{T}d_{m,\bk \ua}\mathcal{T}^{-1}=d_{m,-\bk \da}$ and $\mathcal{T}d_{m,-\bk \da}\mathcal{T}^{-1}=-d_{m,\bk \ua}$. With these unitary transformations the tunneling Hamiltonians of Eq.~\eqref{TunnelingH1} and Eq.~\eqref{TunnelingH2} change to
\begin{subequations}
\begin{align}
H_{T}
&=
\sum_{\bk}
\frac{N_{\text{L}1}}{\pi\nu_{\text{L}}} \
d^{\dag}_{\text{L},\bk \ua}
\gamma_{1}
e^{-i\phi/2}
+
\frac{N_{\text{R}2}}{\pi\nu_{\text{R}}} \
d^{\dag}_{\text{R},\bk \ua}
\gamma_{2}
e^{-i\phi/2}
+\text{H.c.},\label{TunnelingH1_2}
\\
H_{T,\text{ref}}
&=
\sum_{\bk}
\frac{1}{N_{\text{L}1}}
\left(
\Lambda_{\text{L}1}
d^{\dag}_{\text{L},\bk \ua}
+
\Gamma_{\text{L}1}
d^{\dag}_{\text{L},\bk \da}
\right)
\gamma_{1,\text{ref}}
e^{-i\phi_{\text{ref}}/2}
+
\frac{1}{N_{\text{R}2}}
\left(
\Lambda_{\text{R}2}
d^{\dag}_{\text{L},\bk \ua}
+
\Gamma_{\text{R}2}
d^{\dag}_{\text{L},\bk \da}
\right)
\gamma_{2,\text{ref}}
e^{-i\phi_{\text{ref}}/2}
+
\text{H.c.}\label{TunnelingH2_2}
\end{align}
\end{subequations}
Here, we introduced the expressions $\Lambda_{m\ell}\equiv\pi\nu_{m}[
\lambda^{\ua}_{m\ell,\text{ref}}(\lambda^{\ua}_{m\ell})^{*}
+
\lambda^{\da}_{m\ell,\text{ref}}(\lambda^{\da}_{m\ell})^{*}]$
and $
\Gamma_{m\ell}\equiv
\pi\nu_{m}
(\lambda^{\da}_{m\ell,\text{ref}}\lambda^{\ua}_{m\ell}
-
\lambda^{\ua}_{m\ell,\text{ref}}\lambda^{\da}_{m\ell}).$
We notice that the terms $\propto d^{\dag}_{\text{L},\bk \ua}\gamma_{1,\text{ref}}$ and $\propto d^{\dag}_{\text{L},\bk \ua}\gamma_{2,\text{ref}}$ cannot possibly contribute to the Josephson coupling between the two superconducting leads via the TSC islands. On the one hand, these terms cannot yield Cooper pair splitting processes as the MBSs of the four-terminal TSC island couple only to spin-up lead electrons. On the other hand, no direct Cooper pair tunnelling is possible via the reference island due to the conflicting pairing symmetries. By omitting these terms, the tunnelling Hamiltonians of Eq.~\eqref{TunnelingH1_2} and Eq.~\eqref{TunnelingH2_2} further simplify to
\begin{subequations}
\begin{align}
H_{T}
&=
\sum_{\bk}
\frac{N_{\text{L}1}}{\pi\nu_{\text{L}}} \
d^{\dag}_{\text{L},\bk \ua}
\gamma_{1}
e^{-i\phi/2}
+
\frac{N_{\text{R}2}}{\pi\nu_{\text{R}}} \
d^{\dag}_{\text{R},\bk \ua}
\gamma_{2}
e^{-i\phi/2}
+\text{H.c.},\label{TunnelingH1_3}
\\
H_{T,\text{ref}}
&\rightarrow
\sum_{\bk}
\frac{\Gamma_{\text{L}1}}{N_{\text{L}1}} \
d^{\dag}_{\text{L},\bk \ua}
\gamma_{1,\text{ref}}
e^{-i\phi_{\text{ref}}/2}
+
\frac{\Gamma_{\text{R}2}}{N_{\text{R}2}}
d^{\dag}_{\text{L},\bk \ua}
\gamma_{2,\text{ref}}
e^{-i\phi_{\text{ref}}/2}
+
\text{H.c.}\label{TunnelingH2_3}
\end{align}
\end{subequations}
We are now in the position to compute the effective Hamiltonian which implements the rotations around the $z$-axis of the MSQ Bloch sphere. Up to fourth order in the tunnel couplings, the general form of the effective Hamiltonian reads
\begin{equation}
\begin{split}
H_{z,\text{eff}}&=
- P \left(H_{T}+H_{T,\text{ref}}\right) \left(\left[H_{0}+U+U_{\text{ref}}\right]^{-1}\left[1-P\right]\left[H_{T}+H_{T,\text{ref}}\right]\right)^{3}P.
\end{split}
\end{equation}
Here, we have omitted the second order contribution since it only yields a constant energy shift. Furthermore, we have introduced $P=\Pi_{\text{TSC}}\Pi_{\text{SC}}$ where $\Pi_{\text{TSC}}$ is a projector on the charge ground states of the TSC islands and $\Pi_{\text{SC}}$ is a projector on the BCS ground states of the superconducting leads. We now proceed by evaluating the sequences of intermediate states which make up the effective Hamiltonian. In total there are $4!=24$ different ways to transfer a Cooper pair from the left to the right lead. One such process is given by
\begin{equation}
\begin{split}
&\quad\ P
(
d^{\dag}_{\text{R},\bq \ua}
\gamma_{2}
e^{-i\phi/2}
)
(
\gamma_{1}
d_{\text{L},\bk \ua}
e^{i\phi/2}
)
(
d^{\dag}_{\text{R},-\bq \da}
\gamma_{2,\text{ref}}
e^{-i\phi_{\text{ref}}/2}
)
(
\gamma_{1,\text{ref}}
d_{\text{L},-\bk \da}
e^{i\phi_{\text{ref}}/2}
)
P
\\
&=
P
(
d^{\dag}_{\text{R},\bq \ua}
\gamma_{2}
\gamma_{1}
d_{\text{L},\bk \ua}
d^{\dag}_{\text{R},-\bq \da}
\gamma_{2,\text{ref}}
\gamma_{1,\text{ref}}
d_{\text{L},-\bk \da}
)
P
\\
&=
-
\Pi_{\text{TSC}}
(
\gamma_{2}
\gamma_{1}
\gamma_{2,\text{ref}}
\gamma_{1,\text{ref}}
)
\Pi_{\text{TSC}}
\Pi_{\text{SC}}
(
d^{\dag}_{\text{R},\bq \ua}
d^{\dag}_{\text{R},-\bq \da}
d_{\text{L},\bk \ua}
d_{\text{L},-\bk \da}
)
\Pi_{\text{SC}}
\\
&=
e^{i(\varphi_{\text{L}}-\varphi_{\text{R}})}
v_{\bq}
u_{\bq}
u_{\bk}
v_{\bk}
\Pi_{\text{TSC}}
(
\gamma_{2}
\gamma_{1}
\gamma_{2,\text{ref}}
\gamma_{1,\text{ref}}
)
\Pi_{\text{TSC}}
\Pi_{\text{SC}}
(
\gamma_{\text{R},-\bq \da}
\gamma^{\dag}_{\text{R},-\bq \da}
\gamma_{\text{L},\bk \ua}
\gamma^{\dag}_{\text{L},\bk \ua}
)
\Pi_{\text{SC}}
\\
&=
e^{i(\varphi_{\text{L}}-\varphi_{\text{R}})}
v_{\bq}
u_{\bq}
u_{\bk}
v_{\bk}
P
(
\gamma_{2}
\gamma_{1}
\gamma_{2,\text{ref}}
\gamma_{1,\text{ref}}
)
P
\\
&=
(-1)^{n_{0,\text{ref}}}
e^{i(\varphi_{\text{L}}-\varphi_{\text{R}})}
v_{\bq}
u_{\bk}
u_{\bq}
v_{\bk}
P
\hat{z}
P
\end{split}
\label{Sequence1}
\end{equation}
In the fourth
equality we have rewritten the electron operators of the superconducting leads in terms of Bogoliubov quasiparticles,
$d_{m,\bk \ua}=e^{i\varphi_{m}/2}(u_{\bk}\gamma_{m,\bk \ua}+v_{\bk}\gamma^{\dag}_{m,-\bk \da})$ and $d_{m,-\bk \da}=e^{i\varphi_{m}/2}(u_{\bk}\gamma_{m,-\bk \da}-v_{\bk}\gamma^{\dag}_{m,\bk \ua})$. The energy denominator which we pick up in this sequence of intermediate states is given by $1/(E_{\bq}+U)(E_{\bk}+E_{\bq})(E_{\bk}+U)$. There are seven additional sequences which are obtained by appropriately commuting the terms in brackets in the first line of Eq.~\eqref{Sequence1} and which all share the same energy denominators. If we combine them with the corresponding hermitian-conjugated sequences, multiply by the tunneling amplitudes as well as by the energy denominators and carry out the sum over all momenta, we arrive at the the first contribution to the effective Hamiltonian
\begin{equation}
\begin{split}
H'_{z,\text{eff}}&\equiv
\frac{
8
(-1)^{n_{0,\text{ref}}+1}
e^{i(\varphi_{\text{L}}-\varphi_{\text{R}})}
\Gamma_{\text{R}2}
\Gamma_{\text{L}1}^{*}
}{\pi^{2}\nu_{\text{L}}\nu_{\text{R}}}
\sum_{\bk,\bq}
\frac{
v_{\bq}
u_{\bk}
u_{\bq}
v_{\bk}
}
{
(E_{\bq}+U)(E_{\bk}+E_{\bq})(E_{\bk}+U)
} P\hat{z}P
+
\text{H.c.}
\end{split}
\end{equation}
We can now re-express the summation over the momenta in terms of an integral over the density of states. This simplifies the above contribution to
\begin{equation}
H'_{z,\text{eff}} = (-1)^{n_{0,\text{ref}}+1} J_{12}  \cos(\varphi+\varphi_{12}) \hat{z}
\label{Heff_part1}
\end{equation}
where the coupling constant $J_{12}$ and the anomalous phase shift $\varphi_{12}$ were defined in Eq.~(7) of the main text.
\\

Besides the eight sequences discussed above, there are sixteen remaining sequences that come with a different energy denominator. An example of such a sequence of intermediate states is given by
\begin{equation}
\begin{split}
&\quad\ P
(
d^{\dag}_{\text{R},-\bq \da}
\gamma_{2,\text{ref}}
e^{-i\phi_{\text{ref}}/2}
)
(
d^{\dag}_{\text{R},\bq \ua}
\gamma_{2}
e^{-i\phi/2}
)
(
\gamma_{1,\text{ref}}
d_{\text{L},-\bk \da}
e^{i\phi_{\text{ref}}/2}
)
(
\gamma_{1}
d_{\text{L},\bk \ua}
e^{i\phi/2}
)
P.
\end{split}
\label{Sequence2}
\end{equation}
The energy denominator which we pick up for this sequence is given by $1/(E_{\bq}+U)(E_{\bk}+E_{\bq}+2U)(E_{\bk}+U)$. There are fifteen additional sequences which are obtained by appropriately commuting terms in brackets in Eq.~\eqref{Sequence2} and which share the same energy denominator. As before, we can combine all of these sequences with their corresponding hermitian-conjugated counterparts, multiply by the tunneling amplitudes as well as energy denominators and finally carry out the summation over all momenta. This gives the second contribution to the effective Hamiltonian,
\begin{equation}
\begin{split}
H''_{z,\text{eff}} &\equiv
\frac{
16
(-1)^{n_{\text{ref}}+1}
e^{i(\varphi_{\text{L}}-\varphi_{\text{R}})}
\Gamma_{\text{R}2}
\Gamma_{\text{L}1}^{*}
}
{
\pi^{2}\nu_{\text{L}}\nu_{\text{R}}
}
\sum_{\bk,\bq}
\frac{
v_{\bq}
u_{\bk}
u_{\bq}
v_{\bk}
}
{
(E_{\bq}+U)(E_{\bk}+E_{\bq}+2U)(E_{\bk}+U)
} P\hat{z}P
+
\text{H.c.}
\end{split}
\end{equation}
We can again simplify this result by replacing the summation over the momenta with an integral over the density of states.
This gives
\begin{equation}
H''_{z,\text{eff}} = (-1)^{n_{0,\text{ref}}+1} \tilde{J}_{12}  \cos(\varphi+\varphi_{12}) \hat{z},
\label{Heff_part2}
\end{equation}
where the coupling constant $\tilde{J}_{12}$ was defined in Eq.~(7) of the main text.
\\

We can now combine the findings of Eq.~\eqref{Heff_part1} and Eq.~\eqref{Heff_part2} to arrive at the final effective Hamiltonian for rotations around the $z$-axis of the MSQ Bloch sphere,
\begin{equation}
H_{z,\text{eff}} = (-1)^{n_{0,\text{ref}}+1} (J_{12}+\tilde{J}_{12})  \cos(\varphi+\varphi_{12}) \hat{z}.
\end{equation}
\\

The effective Hamiltonian for rotations around the $x$-axis of the MSQ Bloch sphere can be obtained in a very similar way. The only difference is that we choose choose $\lambda^{s}_{\text{L}3}$, $\lambda^{s}_{\text{R}2}$, $\lambda^{s}_{\text{L}1,\text{ref}}$ and $\lambda^{s}_{\text{R}2,\text{ref}}$ as the only non-zero couplings connecting the superconducting leads to the TSC islands. If we thus consistently replace the label $\ell=1$ by $\ell=3$ in the calculations above, we find that
the effective Hamiltonian for the rotations around the $x$-axis of the MSQ is given by
 \begin{equation}
H_{x,\text{eff}} = (-1)^{n_{0,\text{ref}}+1} (J_{32}+\tilde{J}_{32})  \cos(\varphi+\varphi_{32}) \hat{x},
\end{equation}
where the coupling constant $J_{32}$ and the anomalous phase shift $\varphi_{32}$ were defined in Eq.~(7) of the main text.

\section{Effective Hamiltonian for the two-qubit gates}
In this second section of the Supplemental Material, we outline the derivation of the effective Hamiltonian used for the two-qubit entangling operation $XX_{\chi}\equiv\exp(-i\chi \hat{x}_{a}\hat{x}_{b})$ gate for some parameter $\chi$. To derive the effective Hamiltonian of Eq.~(11) in the main text which was used to implement the $XX_{\chi}$-gate, we consistently
carry out the following replacements
\begin{equation}
\begin{split}
\ell=1 \ \rightarrow \ \ell=1,a \quad, \quad
\ell=2 \ \rightarrow \ \ell=2,a  \quad, \quad
\ell=1,\text{ref} \ \rightarrow \ \ell=1,b \quad, \quad
\ell=2,\text{ref} \ \rightarrow \ \ell=2,b
\end{split}
\end{equation}
in our calculations of the previous section on the single-qubit gates, in particular in Eq.~\eqref{Sequence1}. We then indeed reproduce the effective Hamiltonian
\begin{equation}
H_{\text{eff}} =  (J+\tilde{J})  \cos(\varphi+\varphi_{0}) \hat{x}_{a}\hat{x}_{b},
\end{equation}
where the coupling constant $J, \tilde{J}$ and the anomalous phase shift $\varphi_{0}$ were defined in Eq.~(12) of the main text.
\\

As pointed out in Eq.~(14) of the main text, the  $XX_{\chi}$-gate can be used for the implementation of a CNOT and a CNOT$'$ operation, where control and target qubit are exchanged for the CNOT and CNOT$'$ gate. The SWAP operation
which facilitates coherent exchange of quantum information between two four-terminal islands is given by
\begin{equation}
\text{SWAP}= \text{CNOT}'\cdot\text{CNOT}\cdot\text{CNOT}'.
\end{equation}

\end{widetext}

\end{document}